\documentclass[10pt, journal]{IEEEtran}
\usepackage{float}
\usepackage{array}
\usepackage{threeparttable}
\usepackage{cite}
\usepackage{amsmath,amssymb,amsfonts}
\usepackage{algorithmic}
\usepackage{graphicx}
\usepackage{textcomp}
\usepackage{empheq}

\newcommand{\mbb}[1]{\mathbb{#1}}

\newcommand{\m}{\boldsymbol}

\usepackage{lettrine}

\usepackage[colorlinks = true, linkcolor = blue, urlcolor  = blue, citecolor = blue]{hyperref}

\usepackage[subtle]{savetrees}

\title{\textit{\Large \textbf{On Differential Privacy and Traffic State Estimation Problem for Connected Vehicles}}}

\author{Suyash C. Vishnoi, Ahmad F. Taha, Sebastian A. Nugroho, and Christian G. Claudel 
\thanks{Suyash C. Vishnoi and Christian G. Claudel are with the Department of Civil, Architectural, and Environmental Engineering, The University of Texas at Austin, 301 E. Dean Keeton St. Stop C1700, Austin, TX 78712. Ahmad F. Taha is with the Department of Civil and Environmental Engineering, Vanderbilt University, 2201 West End Ave, Nashville, TN 37235. Sebastian A. Nugroho is with the Department of Electrical Engineering and Computer Science, University of Michigan, 1301 Beal Ave., Ann Arbor, MI 48109. Emails:  {\tt\small scvishnoi@utexas.edu, ahmad.taha@vanderbilt.edu, snugroho@umich.edu, christian.claudel@utexas.edu.} This work is partially supported by the National Science Foundation (NSF) under Grants 1636154, 1728629, 1739964, 1917056, 2152928, and 2152450, and USDOT CAMMSE. }}

\begin{document}
\maketitle

\begin{abstract}
This letter focuses on the problem of traffic state estimation for highway networks with junctions in the form of on- and off-ramps while maintaining differential privacy of traffic data. Two types of sensors are considered, fixed sensors such as inductive loop detectors and connected vehicles which provide traffic density and speed data. The celebrated nonlinear second-order Aw-Rascle-Zhang (ARZ) model is utilized to model the traffic dynamics. The model is formulated as a nonlinear state-space difference equation. Sensitivity relations are derived for the given data which are then used to formulate a differentially private mechanism which adds a Gaussian noise to the data to make it differentially private. A Moving Horizon Estimation (MHE) approach is implemented for traffic state estimation using a linearized ARZ model. MHE is compared with Kalman Filter variants namely Extended Kalman Filter, Ensemble Kalman Filter and Unscented Kalman Filter. Several research and engineering questions are formulated and analysis is performed to find corresponding answers. 

\end{abstract}

\section{Introduction and Letter Contributions}

\lettrine{T}{he}   rise of connected vehicles (CVs) technology has provided transportation professionals with additional sources of data to monitor the state of traffic in real time. While more data produces better results when used for state estimation and control, it imposes greater privacy threats on the provider of such data. The location data provided by CVs can be used by criminals for tracking the vehicles, or identifying and profiling the travelers~\cite{chen2013privacy,krumm2009survey}. {Even with sensors that provide aggregate density and speed data, the privacy of individual vehicles is not ensured as it is possible to reconstruct individual trajectories using this data~\cite{le2020differential,xu2017trajectory}.} Rising concern about data privacy in general has led to development of privacy preservation algorithms which can be categorized into anonymity based, obfuscation based and policy based algorithms~\cite{krumm2009survey}. Among these, obfuscation based algorithms such as those which add noise to the data are preferred for tackling location based privacy issues. Such algorithms can be used to ensure differential privacy (DP)~\cite{dwork2014algorithmic} of data. DP is a strong notion of privacy that guarantees the safety of individuals' records when publicly sharing aggregate information from databases. {In context of roadway traffic, DP preserves the location privacy of individual vehicles, both CV and non-CV, when publicly sharing traffic state estimates~\cite{Ny2014,Andre2017}. }

Introducing DP to traffic data however deteriorates the quality of data which could result in a trade off between the level of privacy and estimation accuracy. Since different state estimation algorithms work with different assumptions and approximations, there is reason to believe that some algorithms work better than others when it comes to differentially private state estimation. Therefore it is important to identify such techniques that can produce high quality state estimates while ensuring necessary levels of privacy. 

Past works on differentially private traffic state estimation (TSE) use variants of the Kalman Filter (KF) namely Extended KF (EKF)~\cite{Ny2014} and Ensemble KF (EnKF)~\cite{Andre2017} to perform state estimation. {KFs are known to suffer from certain issues including the absence of state constraints and required assumptions on the distribution of the noise. A technique which traditionally overcomes these drawbacks is Moving Horizon Estimation (MHE)~\cite{rao2001constrained} which is unexplored in the context of differentially private TSE.} Therefore, in this work we implement MHE for  differentilly private TSE and compare its performance with EKF, EnKF and Unscented KF (UKF)~\cite{wan2000unscented}. {Note that unlike \cite{W2} which proposes a privacy preserving MHE to ensure privacy of the estimates produced using non-private data, here we consider that the received data itself is private and use a more traditional MHE formulation. This allows for privacy from the source of data itself.} Also, unlike the past studies which use a first-order traffic model, here we use the second-order Aw-Rascle-Zhang~\cite{aw2000resurrection,zhang2002non} model. Second-order models can reproduce certain real-world traffic phenomena like capacity drop which makes them more suitable for estimation and control purposes. Additionally, we also model junctions which adds more complexity to the model.

Besides, the past work assumes that the speed and density data is obtained from fixed locations on the highway while here we use CVs to obtain data from different parts of the highway. 

{The overall flow of processes in this study is as follows: sensors collect aggregate density and speed data and add privacy preserving noise it. This data is then sent to the network operator who uses it along with a traffic model to perform TSE to obtain density and speed estimates for the road stretch.}

Given that the main research gap on this topic is the absence of a comparative study between different state-estimation techniques for TSE using a second-order model in the presence of DP,
we highlight the main contributions of this letter:
\begin{itemize}    
    \item We present a nonlinear state-space formulation for the second-order ARZ model with junctions. The state-space description is appended to include the measurement model which is also nonlinear. 
    
    \item We derive sensitivity relations for the measured density and speed data. These relations are important for developing differentially private mechanisms that add a Gaussian noise of certain variance to the data to ensure DP. 
   
    \item The performance of various state estimation techniques is investigated in terms of accuracy using the SUMO traffic simulation software in the presence of privacy preserving additive noise. As a departure from estimation based on KFs, we also investigate MHE for TSE. 
\end{itemize}

The letter is organized as follows. Section \ref{s:traffic_model} presents the state-space formulation for the ARZ model and the measurement model. Section \ref{s:DP_MHE} presents the definitions associated with DP, the sensitivity relations for the data and the differentially private mechanism. It also presents the MHE formulation for TSE. Section \ref{s:numerical_study} presents a case study carried out using a realistic traffic simulation software. The letter is concluded by summarizing the results and discussion along with the scope of future work.

\section{Nonlinear ARZ Traffic Dynamics Model}\label{s:traffic_model}
This section presents a state-space formulation for the nonlinear second-order ARZ model~\cite{aw2000resurrection,zhang2002non} describing the evolution of traffic density on highways with ramps. Second-order traffic models, unlike their first-order counterparts, consider traffic density and speed to be independent variables which offers a natural way to incorporate both density and speed data provided by the fixed sensors and CVs. {While other second-order models exist and have been used for TSE in the past~\cite{WANG2022103444}, these models unlike the ARZ model face certain limitations~\cite{daganzo1995requiem} such as physical inconsistency under heterogeneous traffic conditions which makes them unreliable.}

To represent the model as a series of difference, state-space equations, we discretize the ARZ model with respect to both space and time, also referred to as the Godunov scheme~\cite{Godunov1959}. This allows us to divide the highway and the attached ramps into segments of equal length $l$ and time into steps of equal duration $T$. The segments forming the highway are referred to as mainline segments and those forming the ramps are called ramp segments. Throughout the letter, $\Omega, \hat{\Omega},$ and $\check{\Omega}$ denote the set of mainline, on-ramp and off-ramp segments respectively such that $N:=\lvert \Omega\rvert,  N_I:=\lvert \hat{\Omega}\rvert$ and $N_O:=\lvert \check{\Omega}\rvert$. 

The model consists of two states for each segment namely the traffic density (vehicles per unit distance) denoted by $\rho_i[k]$, where $k$ is the index of the time step and $i$ is the index of the segment, and the relative flow (vehicles per unit time) denoted by $\psi_i[k]$. The discrete time traffic density and relative flow conservation equations for any Segment $i\in\Omega$ can be written for any time step $k$ as
\begin{subequations}\label{e:Godunov_update_equations1}
\begin{align}
 \small   \rho_i[k+1]&=\rho_i[k]+\frac{T}{l}(q_{i-1}[k]-q_i[k]),\\\label{e:Godunov_update_equations2}
    \psi_i[k+1]&= \frac{\tau-1}{\tau} \psi_i[k]+\frac{T}{l}(\phi_{i-1}[k]-\phi_i[k])+\frac{v_f}{\tau}\rho_i[k].
\end{align}
\end{subequations}
Here, $q_i[k]$ and $\phi_i[k]$ denote the quantities traffic flow and relative flux leaving Segment $i\in\Omega$ at time step $k$, and $v_f$ denoting the free flow speed of a segment, and $\tau$ are parameters of the ARZ model.  Similar equations can be written for ramp segments as well. Mathematical expressions for $q_i[k]$ and $\phi_i[k]$ can be written using the expressions for certain other quantities namely the demand ($D[k]$), supply and driver characteristic ($w[k]$) which are not presented in this article for brevity. 

These quantities are given as nonlinear functions of the states and inputs defined similar to \cite{kolb2018pareto}. 
The state vector for this system can be defined as 
$$ \m x[k] :=[\rho_i[k]\hspace{1mm} \psi_i[k]\hspace{1mm} \ldots\hspace{1mm}\hat{\rho_j}[k]\hspace{1mm} \hat{\psi_j}[k]\hspace{1mm} \ldots\hspace{1mm}\check{\rho_l}[k]\hspace{1mm} \check{\psi_l}[k]\hspace{1mm} \ldots]^{\top}
$$
where $\m x[k] \in \mbb{R}^{2(N+N_I+N_O)}$ and $i\in\Omega$, $j\in\hat{\Omega}$ and $k\in\check{\Omega}$. The variables with $\hat{\cdot}$ are associated with the on-ramps and those with $\check{\cdot}$ are associated with off-ramps. The input vector is defined as,
\begin{align*}
   \m u[k] := [&D_{in}[k] \hspace{1mm}w_{in}[k]\hspace{1mm}\rho_{out}[k] \hspace{1mm} \ldots\hspace{1mm} \hat{D}_{in,j}[k]  \hspace{1mm} \hat{w}_{in,j}[k]  \ldots\hspace{1mm}\hspace{1mm} \check{\rho}_{out,l}[k]\hspace{1mm}\ldots]^{\top}
\end{align*}
where $\m u[k] \in \mbb{R}^{3+2N_I+N_O}$, $j\in\hat{\Omega}$ and $l\in\check{\Omega}$.

The evolution of traffic density and relative flow
 described in \eqref{e:Godunov_update_equations1} can be written in a compact state-space form as follows 

\begin{empheq}[box=\fbox]{align}
    \m x[k+1]=\m A\m x[k]+\m G\m f(\m x,\m u),
\label{eq:state_space_gen}
\end{empheq}

\noindent where $\m A \in \mbb{R}^{n_x\times n_x}$ for $n_x := 2(N+N_I+N_O)$ represents the linear portion of the dynamics of the system, $\m f:  \mbb{R}^{n_x}\times  \mbb{R}^{n_u}\rightarrow  \mbb{R}^{n_x}$ where $n_u=3+2N_I +N_O$ is a vector valued function representing nonlinearities in the state-space equation, and $\m G \in \mbb{R}^{n_x\times n_x}$ is a matrix representing the distribution of nonlinearities. The nonlinearities in $\m f$ are in the form of minimum of weighted nonlinear functions of the states and inputs. The structure of the above mentioned functions are similar to those presented in \cite{kolb2018pareto}. {The modeling approach can be applied to roads with any number of lanes given the maximum density is adjusted based on the number of lanes.} Next, we discuss the measurement model which is also nonlinear in nature.

{We consider two types of sensors, fixed sensors like the inductive loop detectors and CVs. This study assumes that it is possible to retrieve aggregate density and speed data for road segments from both these sensors. Such data can be obtained from fixed sensors directly using techniques such as in \cite{lee2011density}. With CVs, the average speed of a segment is assumed to be the average of the speed of all the queried CVs in a segment similar to~\cite{bekiaris2017highway}. To obtain density data from CVs, we assume additional functionality like spacing measurement equipment available in advanced driver assistance systems~\cite{seo2015estimation} or availability of vehicular ad-hoc networks (VANETs) which allow vehicles to communicate with each other in a neighbourhood around the queried CV~\cite{panichpapiboon2008evaluation}. A sufficient penetration of CVs is necessary on the segments which are queried for data. The spacing data or neighbourhood counts can then be converted to density measurements before adding the privacy preserving noise to them and sending them to a network operator to perform estimation.}

Among these measurements, density $\rho_i[k]$ for any mainline segment $i\in\Omega$, and similarly for the ramps, is directly a state and is used as it is, while the speed $v_i[k]$ can be written in terms of the states as follows:
\begin{align}
    v_i[k] = \frac{\psi_i[k]}{\rho_i[k]} - p(\rho_i[k]),\nonumber
\end{align}
where $p(\cdot)$ is called the pressure function and is defined as part of the ARZ model framework.
We define a nonlinear measurement function $\m h(\m x[k])$ such that
\vspace{-3mm}
\begin{align}
    h_{2i-1}(\m x[k]) = x_{2i-1}[k], \textrm{ and }
    h_{2i}(\m x[k]) = \frac{x_{2i}[k]}{x_{2i-1}[k]} - p(x_{2i-1}[k]).\nonumber
\end{align}
Now, we can define the measurement vector $\m y[k]$ as 
\begin{empheq}[box=\fbox]{align}
\label{e:meas_model}
    \m y[k] = \m C[k]\m h(\m x[k])+\m \nu[k],
\end{empheq}
where $\m C[k]$ is the observation matrix at time $k$ describing the availability of measurements from sensors. Note, that the observation matrix here is variable in time because of the measurements from CVs which are taken from different segments at different times. At any time $k$, $n_p[k]$ is the number of measurements. Here, $\m \nu[k]\in\mathbb{R}^{n_{\nu}[k]},n_{\nu}[k]=n_p[k]$ lumps all the measurement errors including the sensor noise into a single vector.

In the following section we discuss some definitions related to differential privacy with respect to the traffic data, the dynamics \eqref{eq:state_space_gen}, and the measurement model \eqref{e:meas_model}.

\section{Differential Privacy of Traffic Data and MHE}\label{s:DP_MHE}
Making the data differentially private is considered as an adequate measure against privacy attacks such as unwanted tracking of vehicles and identifying individuals based on location data. {While certain cryptographic methods maintain privacy by preventing attackers from reading the data, under the possibility that the attacker finds a way to read it, differential privacy adds another layer of defense which statistically guarantees that individual's records cannot be extracted from the data set. It also allows sharing of estimates obtained from this data with third-parties keeping the same guarantee.} DP is achieved by processing the data through differentially private mechanisms which are functions that take entire data sets as input and produce a differentially private output. In the following sections we discuss some definitions that are needed to formally define DP.
\vspace{-1mm}
\subsection{Adjacency and DP}
DP is defined in terms of adjacent data sets. Mathematically, \textit{adjacency} is defined as a binary symmetric relation denoted by $Adj$ on a space of data sets, say $D$, such that for $d,d'\in D$ $Adj(d,d')$ holds if and only if $d$ and $d'$ differ by the data of a single individual. In this work, we consider two spaces of data sets, the traffic density data sets and the traffic speed data sets which are composed of the average vehicle density and average vehicle speed values from several road segments and several time steps. Two data sets from either of these spaces are said to be adjacent if they differ by the trajectory of a single vehicle. 
With this definition of adjacency, a differentially private mechanism can be defined similar to \cite{Andre2017} as follows.

\vspace{0.2cm}

\noindent \textbf{Definition 1.} Let $D$ be a space of data sets, and let $(\textrm{R},\mathcal{M})$ be a measurable space where $\mathcal{M}$ is a $\sigma$-algebra on $\textrm{R}$. Let $\epsilon,\delta \ge 0$. A mechanism $M:D \rightarrow R$ is $(\epsilon,\delta)-$differentially private if for all $d,d'\in D$ such that $\mathrm{Adj}(d,d')$, we have
\vspace{-1mm}
\begin{equation}
    \label{e:differential_privacy}
    \mathbb{P}(M(d)\in S)\le e^{\epsilon}\mathbb{P}(M(d')\in S)+\delta, \forall S \in \mathcal{M}
\end{equation}    

This means that the distribution of the outputs produced by the mechanism $M$ on any two adjacent data sets is very close which makes it difficult to determine which data set was used as input by looking at the output of the mechanism. Thus, attackers are unable to extract individual specific information from the mechanism's output. Releasing this output instead of the original data protects individual's privacy against attacks. $M$ is therefore said to provide $(\epsilon,\delta)$-DP to the data. Smaller values of both $\epsilon$ and $\delta$ provide higher privacy. In this work, we assume that such a privacy preserving mechanism is applied to the density and speed data collected by the fixed sensors and CVs at the source and the output is sent to the network operator.

An important property~\cite{Ny2014} which allows the network operator to use this data for state estimation and control while maintaining the DP guarantee is called \textit{resilience-to-post-processing}. According to this property, if another mechanism is applied to the output of a differentially private mechanism, the obtained result will have the same DP guarantees as the initial output. In context of this work, the mechanism applied after receiving the differentially private outputs from the sensors is the estimation process. Thus, the final state estimates are also differentially private. 

To write the mechanisms capable of producing differentially private outputs, we need to first define sensitivity relations for the two types of data sets.

\subsection{Sensitivity relations}
The sensitivity of a function is defined as the maximum difference in the value of the function produced by two adjacent data sets. In this work we are concerned about the sensitivity of data coming from the traffic sensors. {Since both the type of sensors considered in this study provide the same two type of data, that is the segment density and speed, we do not have a separate sensitivity relation for CV data than for fixed sensor data.} Specifically, we care about the Euclidean norm between adjacent data sets, that is, $\Vert \m \rho-\tilde{\m \rho}\Vert_2$ and $\Vert \textbf{v} - \tilde{\textbf{v}}\Vert_2$ where $\m \rho, \tilde{\m \rho}$ and $\m v, \tilde{\m v}$ are any two adjacent pairs of density and speed data respectively. We can write
\vspace{-1mm}
\begin{align*}
    \Vert \m \rho-\tilde{\m \rho}\Vert^2_2 = \sum_{k=0}^{\infty}\sum_{i=1}^{n_p[k]}|\rho^i[k]-\tilde{\rho}^i[k]|^2
\end{align*}
where $\rho^i[k]$ represents the density measured at the $i^{th}$ density sensor at time step $k$. Largest sensitivity value occurs when the differentiating vehicle passes all the sensors at different times in the two data sets. {Since the density of a segment can be defined as the number of vehicles per unit length of the segment, the density measurements in the two data sets can be assumed to differ by $\dfrac{1}{l}$ when the differentiating vehicle is present on a measured segment in one data set and absent in the other as the difference is caused by a single vehicle being present or absent on that segment. The total time during which the density for a segment differs between the two data sets at  any such instance can be approximated based on the average time spent by a vehicle on that segment. Let this average time be denoted by $T_{avg}$, which can be approximated using past CV data for that stretch or by using a simulation-based approach as in \cite{Andre2017}. Here $T_{avg}$ for all segments is assumed to be the same but in practice a different $T_{avg}$ can be computed for different segments to get a better approximation of the sensitivity.} At all other times the measured densities would be the same in both the data sets. Then,
\begin{align}\label{e:dprho1}
    \Vert \m \rho-\tilde{\m \rho}\Vert^2_2 =\sum_{k=0}^{\infty}\sum_{i=1}^{n_p[k]}|\rho^i[k]-\tilde{\rho}^i[k]|^2\le \sum_{i=1}^{N_p}2T_{avg}\left(\dfrac{1}{l}\right)^2,
\end{align}
\vspace{-1mm}
\begin{align}\label{e:dprho2}
    \implies\Vert \m \rho-\tilde{\m \rho}\Vert_2 \le \dfrac{1}{l}\sqrt{2N_pT_{avg}}=:\Delta_{\rho},
\end{align}
where $N_p$ is the maximum number of sensors on the highway stretch at any time and $\Delta_{\rho}$ is the sensitivity of the density data sets. 

Similarly, for speed measurements we can write
\vspace{-1mm}
\begin{align}
    \label{e:DP_vel}
    \Vert \m v-\tilde{\m v}\Vert^2_2 = \sum_{k=0}^{\infty}\sum_{i=1}^{n_p[k]}|v^i[k]-\tilde{v}^i[k]|.
\end{align}
The effect of the absence or presence of a single vehicle in the segment on the average speed of that segment can be approximately captured indirectly with the help of the equilibrium speed-density relationship of the ARZ model~\cite{aw2000resurrection,zhang2002non} given as
\begin{align}\label{e:pressure_equilibrium_relationship}
    V_e(\rho)=v_f\left(1-\left(\frac{\rho}{\rho_m}\right)^\gamma\right),
\end{align}
which relates the equilibrium speed $V_e$ of a road segment with the density of that segment. Here, $\rho_m$ denoting the maximum density of a segment, and $\gamma$ are parameters of the ARZ model. We can replace the speeds in the right hand side of \eqref{e:DP_vel} with the expression in \eqref{e:pressure_equilibrium_relationship} with $\gamma=1$ and simplify it to get
\vspace{-2mm}
\begin{align}
    |v^i[k]-\tilde{v}^i[k]| = \left|\dfrac{v_f}{\rho_m}\left(\rho^i[k]-\tilde{\rho}^i[k]\right)\right|.
    \label{e:presense}
\end{align}
Then using the same idea as for the density, we can write 
\begin{align}
    \label{e:dpvel2}
    \Vert \m v-\tilde{\m v}\Vert_2 \le \dfrac{v_f}{\rho_ml}\sqrt{2N_pT_{avg}}=:\Delta_{v},
\end{align}
where $\Delta_{v}$ is the sensitivity of the speed data sets. Here, $\gamma=1$ is chosen arbitrarily to simplify the expression \eqref{e:presense} to a known constant value. Though $\gamma$ varies between 1 and 2~\cite{kolb2018pareto} \eqref{e:dpvel2} serves as a good upper bound in most cases since $\rho$ lies between quarter to one-third of $\rho_m$ under normal flow conditions. The sensitivity $\Delta_v$ can be modified under specific scenarios using empirical tests or using a traffic simulation approach as in~\cite{Andre2017}. {In general, ensuring a realistic value of the sensitivity avoids a large privacy-utility trade off.} 
\vspace{-2mm}
\subsection{Differentially private mechanisms}\label{s:mechanism}
Using the sensitivity relations from the previous section, we can implement a Gaussian Mechanism~\cite{Ny2014} which ensures $(\epsilon,\delta)-$DP. 

Let $K=Q^{-1}(\delta)$ for $Q(x)=\dfrac{1}{\sqrt{2\pi}}\int_x^{\infty}e^{-u^2/2}du$, and $\kappa_{\delta,\epsilon}=(K+\sqrt{K^2+2\epsilon})/(2\epsilon)$, then a mechanism publishing the sequence $\bar{\rho} = \rho + w_{\rho}$ where $w_{\rho}$ are zero mean iid Gaussian random variables with variance $\kappa_{\delta,\epsilon}^2\Delta_{\rho}$ is $(\epsilon,\delta)$-differentially private. Here $\rho$ is the measured density data and $\bar{\rho}$ is the differentially private output produced by the mechanism which will be sent to the network operator.

Similarly, a mechanism publishing the sequence $\bar{v} = v + w_{v}$ where $w_{v}$ are zero mean iid Gaussian random variables with variance $\kappa_{\delta,\epsilon}^2\Delta_{v}$ is also $(\epsilon,\delta)$-differentially private. Here $v$ is the measured speed data and $\bar{v}$ is the differentially private output of the mechanism. For the mechanisms defined here, the output itself is a data set which will henceforth be called differentially private data. In the next section, we discuss the MHE approach applied for TSE using the differentially private data produced by the mechanisms.

\vspace{-2mm}
\subsection{Moving horizon estimator under DP}\label{sec:state_estimation_mhe}

The objective of this article is to investigate the TSE performance using ARZ model when considering differentially private data coming from the fixed sensors and CVs. To do so, here we implement a linear MHE approach using linearized versions of the process and measurement models obtained using a first-order Taylor series approximation. Throughout this section, $N$ denotes the size of the horizon for optimization.
\subsubsection{Decision variables and objective function}
The decision variables for the optimization problem solved at time step $k$ are the state vectors from time step $k-N$ to $k$ denoted by $\m x_k[t]\hspace{1mm}\forall\hspace{1mm} t\in[k-N,k]$. From the obtained solution we set the final value of the vector $ \hat{\m x}_k[k] = \m x_k[k]$. The objective function at time step $k\in[N+1,\infty]$ is denoted by $J[k]:=J$ and is given as

\begin{align}
    \label{e:mhe_objective}
   \small J=& \hspace{1mm}\mu||{\m x}_k[k-N]- \bar{\m x}[k-N]||^2\nonumber+w_1\hspace{-2mm}\sum^k_{i=k-N}\hspace{-2mm}|| \m y[i]-( \tilde{\m C}_{i}\m x_k[i]+\m c_{2i}) ||^2\nonumber\\
    &+w_2\hspace{-2mm}\sum^{k-1}_{i=k-N}\hspace{-2mm}|| \m x_k[i+1]\hspace{-0.5mm}-\hspace{-0.5mm} (\tilde{\m A}_{i}\m x_k[i] + \m B_{i}\m u[i] + \m c_{1i}) ||^2.
\end{align}

Here, $\bar{\m x}[k-N]$ is a prediction of $\m x[k-N]$ based on a previously obtained state estimate and is expressed as
\begin{align}\label{e:prediction}
    \bar{\m x}[k\hspace{-0.5mm}-\hspace{-0.5mm}N] = \m A \hat{\m x}[k\hspace{-0.5mm}-\hspace{-0.5mm}N\hspace{-0.5mm}-\hspace{-0.5mm}1] + \m G\m f(\hat{\m x}[k\hspace{-0.5mm}-\hspace{-0.5mm}N\hspace{-0.5mm}-\hspace{-0.5mm}1],\m u[k\hspace{-0.5mm}-\hspace{-0.5mm}N\hspace{-0.5mm}-\hspace{-0.5mm}1]).
\end{align}
 
The notation $\m y[i]$ defines the data vector at time $i\in [k-N,k]$, $\tilde{\m A}_{i}, \m B_{i}$ and $\m c_{1_i}$ are parameters of the linearized state-space equation $\hspace{0.5mm}\forall \hspace{0.5mm}i \in[k-N,k-1]$, and $\tilde{\m C}_{i}$ and $\m c_{2_i} $ are parameters of the linearized measurement model $\hspace{0.5mm}\forall \hspace{0.5mm}i \in [k-N,k]$. Here, $\tilde{\m A}_{k}, \m B_{k}$ and $\m c_{1_k}$ are computed at $(\m x_o,\m u[k])$ where $\m x_o\hspace{-1mm}=\hspace{-1mm} \sum_{i=k-1-N}^{k-1} \m x_{k-1}[i]/(N+1)$, $\tilde{\m C}_{k}$ and $\m c_{2_k} $ are computed at $\m x_o$.

\subsubsection{Constraints and optimization problem}
The problem only consists of the upper and lower bounds on state values as follows
\begin{align}\label{e:bounds}
    \m x_{\mathrm{min}} \le \m x_k[i] \le \m x_{\mathrm{max}}, \forall \hspace{1mm}i \in [k-N,k]
    \vspace{-3mm}
\end{align}
where $\m x_{\mathrm{min}} = \vec{0}$, and $\m x_{\mathrm{max}}=[\rho_m \hspace{2mm} \rho_m v_f \hspace{2mm}\rho_m \hspace{2mm}\rho_m v_f \hspace{2mm}\cdots \hspace{2mm}\rho_m \hspace{2mm}\rho_m v_f]^T$. The above objective and constraints are used to write the following optimization problem
\begin{align}\label{e:optimization}
\underset{\m x_k[k-N],\dots,\m x_k[k]}{\textrm{minimize}} &&J[k],\;\;\; \;\; \textrm{subject to}&&\eqref{e:bounds}.
\end{align}
The objective function $J[k]$ can also be expressed as a sum of quadratic and linear terms of the state vectors. Defining $\m z_k$ by concatenating the decision variables from \eqref{e:optimization} such that $\m z_k = [\m x_k[k-N]^T\hspace{1mm} \m x_k[k-N+1]^T\hspace{1mm} \cdots\hspace{1mm} \m x_k[k]]^{T}$, we can write the optimization problem \eqref{e:optimization} in standard form as follows
\vspace{-1mm}
\begin{empheq}[box=\fbox]{align}
\label{e:QP}
\small \underset{\m z_k}{\textrm{minimize}} &\; \; \m z_k^T\m H\m z_k+\m q^T\m z_k, \;\;\; \textrm{subject to} \; \; \m z_{\mathrm{min}}\le \m z_k \le \m z_{\mathrm{max}}
\end{empheq}
where $\m H\in\mathbb{R}^{(N+1)n_x\times (N+1)n_x}$ and $\m q\in\mathbb{R}^{(N+1)n_x}$ consist of the coefficients of the quadratic and linear terms in the objective respectively. $\m z_{\mathrm{min}}=[(x_{min}^T)_{\times (N+1)}]^T$ and $\m z_{\mathrm{max}}=[(x_{max}^T)_{\times (N+1)}]^T$ denote bounds on $\m z_k$. It can be shown that $\m H$ is a positive definite matrix which makes \eqref{e:QP} a convex quadratic program (QP) that can be solved efficiently using readily available QP solvers like CPLEX or MATLAB's \texttt{quadprog} function. Hence, problem~\eqref{e:QP} is computationally tractable.

\section{Case Studies using SUMO}\label{s:numerical_study}

In this section, we apply the implemented MHE along with EKF, UKF and EnKF, on a traffic simulation example generated in SUMO which is an open source the traffic micro-simulation software to compare their performance while keeping the data differentially private. 
All the simulations are carried out using MATLAB R2019b running on a 64-bit Windows 10 with 3.6GHz Intel$^\textrm{R}$ Core$^\textrm{TM}$ i7-7700 CPU and 65GB of RAM. We use the \texttt{quadprog} function of MATLAB to solve the MHE optimization problem. 

The main idea of this case study is to test the performance of the state estimation techniques under different conditions of privacy. In particular, we are interested in knowing the answers to the following questions:
\begin{itemize}
    \item \textit{Q1:} How does the number of CV-segments impact the state estimation performance of each technique while ensuring DP of data? 
    \item \textit{Q2:} What is the impact of the level of privacy on the state estimation performance of each technique? 
\end{itemize}
\vspace{-3mm}
\subsection{Highway and sensor setup}\label{s:highway_structure}
In this study, we model a highway stretch of length 1.5 km with two on-ramps at 0.3 and 0.9 km from the start and two off-ramps at 0.6 and 1.2 km from the start. Additional 100 m segments are modeled in SUMO before all the entry points and following all the exit points of the highway whose data serves as input for the system. We use the Weidemann 99 car-following model with default parameters.
The ARZ model parameters are calibrated using simulated data from SUMO. The selected values are $v = 102$ km/hr, $\rho_m = 333$ veh/km, $\tau = 60,$ and $\gamma = 2$. Under the Godunov scheme, the highway and ramps are divided into segments of length 100 m each with a time-step value of 1 s, which satisfies the CFL condition. {The segment mean speeds are provided directly by SUMO while the segment densities can be computed from the vehicle count data provided by SUMO for each segment.} There are a total of 38 states in this highway system consisting of 15 mainline segments and 4 ramps. {Since the local state-space dynamics for segment-type combinations as in \eqref{e:Godunov_update_equations1} are the same irrespective of the overall structure, the state estimation performance here should be representative of the performance in general.} 

{We force a congestion on the highway to create an interesting scenario for comparison of TSE methods. The mainline demand is kept as 2050 veh/hr throughout except between 200-400 sec when it is increased to 6050 veh/hr owing to say a rush hour. The on-ramp demands are kept as 320 veh/hr and 300 veh/hr respectively. A variable speed sign is implemented in SUMO to emulate a situation where an accident has occurred on Segment 11 of the highway mainline. The maximum allowed speed for this segment is artificially reduced to 10.08 km/hr between 200-400 sec and 50.95 km/hr between 400-500 sec. }

Throughout the case study, the fixed sensors are assumed to be placed on the output segments of the network which is necessary to make the system observable. {We assumes that there is a sufficient number of CVs on the highway to obtain the density and speed values of decent quality from any road segment. We also assume that we can only query a limited number of CVs at a time due to bandwidth constraints. At every time step we select a subset of segments to obtain data from. In the case study, we select a set of segments at the beginning and update it after every four time steps. At every update, the current segments in the set are replaced by segments right ahead of them. The last mainline segment is replaced by the first mainline segment. Note that a better method to select CVs for querying may be available but is not explored here.}
{No measurement noise is added to the data apart from the privacy preserving noise.}

\vspace{-2mm}
\subsection{Implementation of estimation techniques}
\vspace{-1mm}
\paragraph{Parameter tuning}
In this work, for all the KF variants, we use diagonal process and measurement noise co-variance matrices of the form $\m Q = q\m I$ and $\m R = r\m I$ where $q,r\in\mathbb{R}_+$ and $\m I$ in each case is an identity matrix of appropriate dimensions. The initial guess for the estimate noise co-variance matrix is taken as $\m P=10^{-3}\m I$. We manually tune $q$ and $r$ for different arrangements of sensors and different privacy levels based on the minimization of the root mean squared error ($\mathrm{RMSE}$) of estimated states. The weights in the MHE objective function are also similarly tuned. Regarding other parameters, for UKF~\cite{wan2000unscented}, we set the following values: $\alpha=0.1, \kappa=-4$ and $\beta=2$, for EnKF~\cite{evensen2003ensemble}, we set the number of ensemble points to $100$, and for MHE, we set $N$ to $10$. These values are found to be sufficient for the respective techniques except wherever specified.

\paragraph{External bounds in KF}
The KFs can produce negative values of the states which are not allowed in the process model \eqref{eq:state_space_gen}. It results in numerical issues and forces the estimation to stop. To avoid this estimates are projected to within physical bounds. In case of UKF, the sigma points need to be individually bounded with a lower bound greater than zero to avoid numerical issues within UKF. {This method of projecting vectors for EKF and UKF has been shown to fit in the KF theory mathematically and is among the popular methods mentioned in \cite{simon2010kalman}.}

\paragraph{Choice of comparison metrics}
Parameter tuning is done using the $\mathrm{RMSE}$ of the estimated states. However, since the relative flow does not hold a direct significance for professionals, we chose to compare the techniques based on the $\mathrm{RMSE}$ of density and speed which have more general value.  

\subsection{Results and discussion}\label{s:results}
\subsubsection{Impact of number of CV-segments}
\begin{figure}
\centering
    \includegraphics[width=0.5\textwidth]{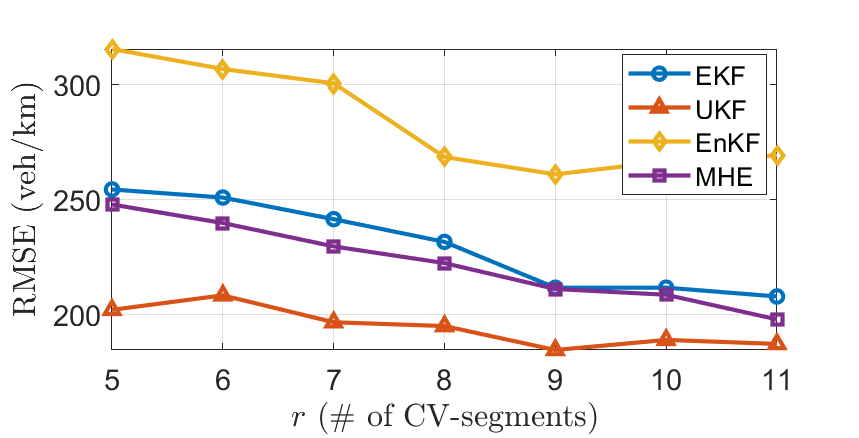}  
    \includegraphics[width=0.5\textwidth]{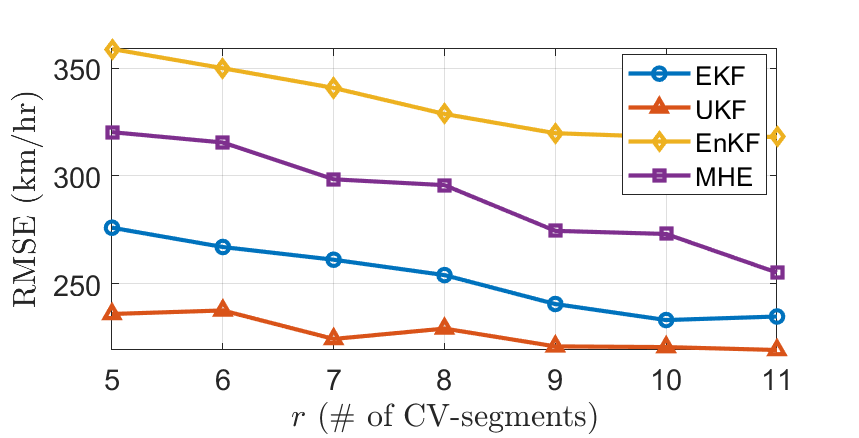}    
    \caption{$\mathrm{RMSE}$ for (a) density [top] and (b) speed [bottom] with increasing number of CV-segments while considering (1,0.05)-DP.}
    \label{f:num_CVs_ds}
\end{figure}

 Here, we test the impact of increasing the number CV-segments on the performance of different state estimation techniques. We vary the number of CV-segments from 5 to 11 while keeping them as far apart as possible. Exact arrangement is omitted for brevity. Privacy preserving noise is added to the measurement values based on the mechanism in Section \ref{s:mechanism} to make the data $(1,0.05)$-differentially private. Fig. \ref{f:num_CVs_ds} presents the plot of $\mathrm{RMSE}$ for the estimated density and speed for each of the techniques. {The computation time per time step of simulation for EKF, UKF, EnKF, and MHE are 0.06, 0.026, 0.040, and 0.045 seconds respectively.} These include the time taken from receiving the data to producing the estimate for one time step.

\begin{figure}
    \centering
    \includegraphics[width=0.4\textwidth]{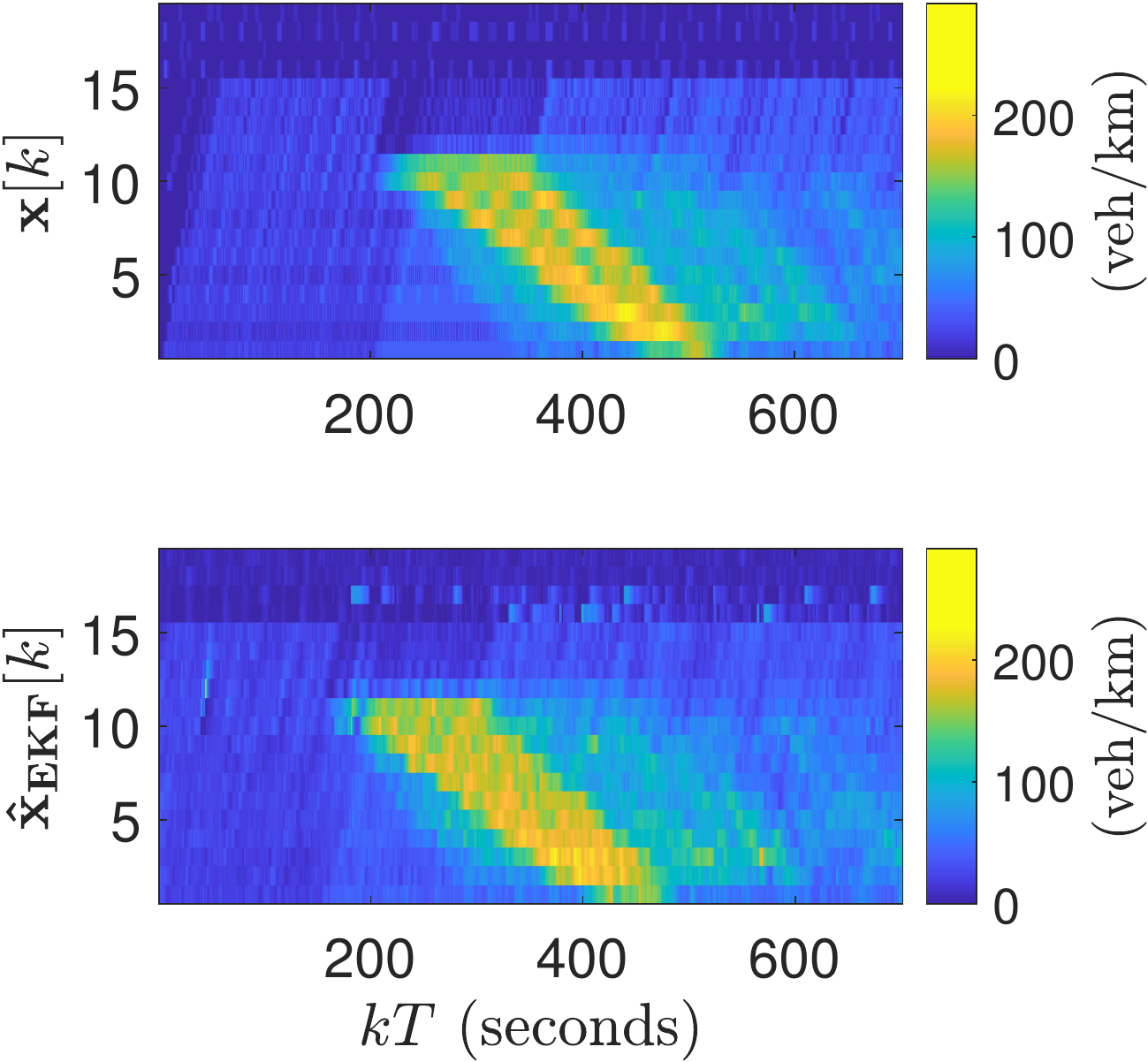} \includegraphics[width=0.4\textwidth]{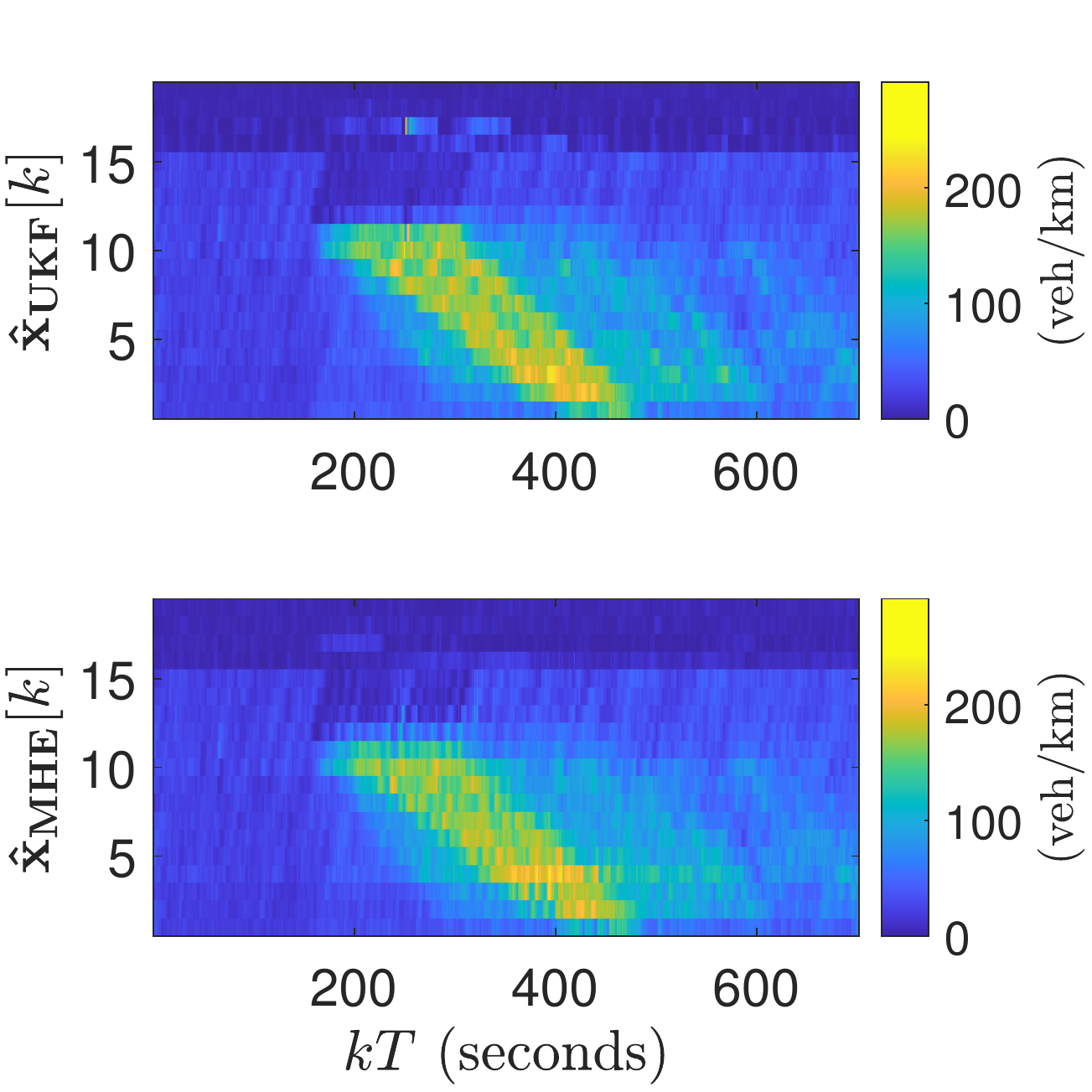}
    \caption{Traffic densities $\textbf{x}[k]$ obtained from SUMO along with estimates from EKF denoted by $\hat{\m x}_{\mathrm{EKF}}[k]$, UKF denoted by $\hat{\m x}_{\mathrm{UKF}}[k]$, and MHE denoted by $\hat{\m x}_{\mathrm{MHE}}[k]$ while considering $(1,0.05)-$DP and 7 CV-segments evenly spaced on the highway stretch.}
    \label{f:states}
\end{figure}

It is observed that the estimation performance for all the techniques improves with increasing number of CV-segments which is expected. EnKF sometimes has more variation in consecutive $\mathrm{RMSE}$ values as compared to other techniques which can be attributed to the associated randomness. Overall, EnKF performs the worst while UKF performs the best, closely followed by both MHE and EKF in case of density estimation and EKF in case of speed estimation. It is interesting that MHE falls behind EKF in case of speed estimation. This comparison in performance is also observed in the following tests. This is mentioned here to avoid repetition later. Fig. \ref{f:states} presents a plot of the actual versus estimated density values obtained using UKF, EKF and MHE.

\subsubsection{Impact of DP parameters}
\begin{figure}
    \centering
    \includegraphics[width=
    0.50\textwidth]{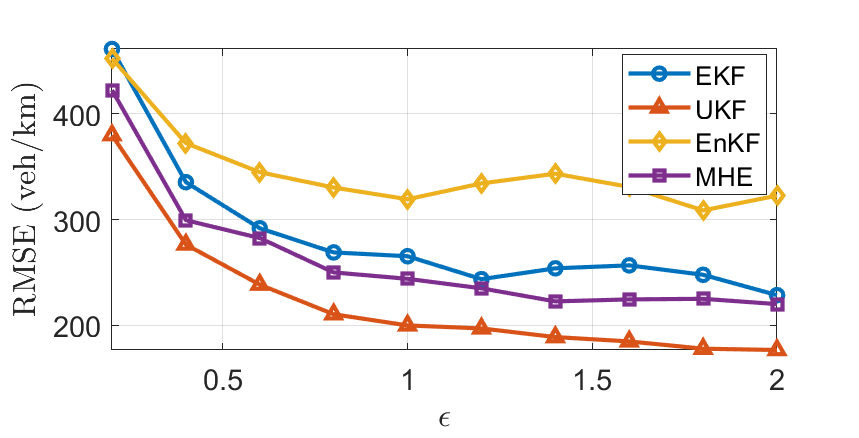}
    \caption{$\mathrm{RMSE}$ for densities with changing $\epsilon$ while keeping $\delta=0.05$.}
    \label{f:epsilon_density}
\end{figure}

\begin{figure}
 \centering   \includegraphics[width=0.50\textwidth]{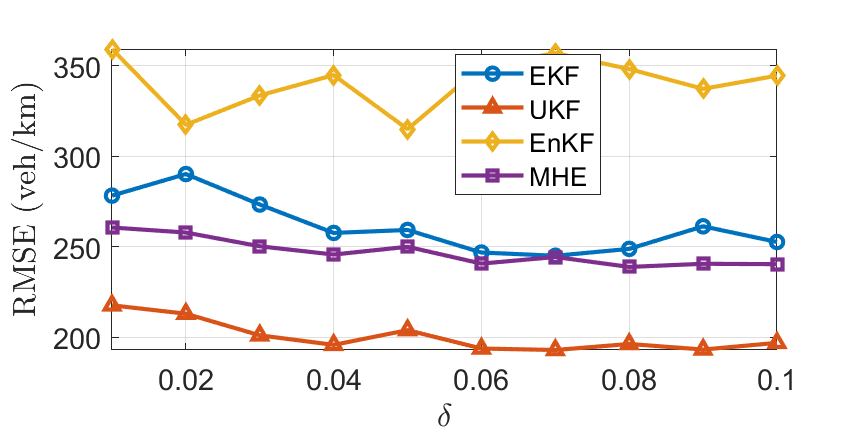}   
    \caption{$\mathrm{RMSE}$ for densities with changing $\delta$ while keeping $\epsilon=1$.}
    \label{f:delta_density}
\end{figure}

Privacy in this study depends on two parameters $\epsilon$ and $\delta$ which have their own significance in the DP definition. In this section, we test the impact of varying these parameters on the state estimation performances. We keep the same configuration of fixed sensors as in the previous section while the number of CV-segments is fixed to 5. Fig. \ref{f:epsilon_density} and Fig. \ref{f:delta_density} present the variation in $\mathrm{RMSE}$ values for density for each technique with changing epsilon keeping a constant $\delta=0.05$, and changing delta keeping $\epsilon= 1$ respectively. The plots for speed in both cases are very similar to density and are omitted for brevity. The co-variance matrices are tuned as necessary to obtain the best performance.

All the techniques show a similar variation in performance with respect to privacy changes. It is observed that the impact of $\epsilon$ is more profound than that of $\delta$ when both are varied between respective reasonable bounds. Specifically, the variation in performance is small over the full range of selected $\delta$ values. On the other hand, the variation is small for $\epsilon$ values above 1, but the performance quickly worsens as we approach 0. While more research might be needed under various scenarios, from the obtained results it can be stated that it is possible to increase the level of privacy to a certain extent without worrying about much additional degradation of estimation quality. Beyond that point, a trade-off would be more apparent and should be considered more seriously. 

\subsubsection{Discussions and preliminary answers} We provide some preliminary suggestions regarding the questions posed earlier in this section:
\vspace{-1mm}
\begin{itemize}
    \item \textit{A1:} State estimation error decreases with an increase in the number of CV-segments. 
    UKF outperforms the other methods in both density and speed estimation while EnKF's performance is the worst. EKF and MHE perform comparably.
    
    \item \textit{A2:} All the techniques show similar variation in performance with change in privacy levels. In general, $\epsilon$ has more influence on the estimation quality than $\delta$.
    
\end{itemize}
{A drawback of the present study is that it assumes that both the CVs and fixed sensors provide the same measurement values for a segment if present simultaneously. This may not always be true and a reliable approach for data integration may be needed. Studying the data integration problem considering different aggregate measurements from sensors or using trajectory data from CVs for state estimation and its impact on privacy are possible future directions of work.}
{Also, while not studied in this work, the advantage of MHE in implementing arbitrary relations between states which are otherwise un-modeled in the dynamics can also be explored. }
\vspace{-3mm}
\bibliographystyle{IEEEtran}	\bibliography{biblio}

\end{document}